\documentclass[pra,twocolumn,amsmath,amssymb]{revtex4-1} %

\usepackage{epsfig,amsmath}
\usepackage{subfigure}
\usepackage{graphicx}
\usepackage{dcolumn}
\usepackage{stmaryrd}
\usepackage{mathrsfs}
\usepackage{pifont}
\usepackage{amsthm}
\usepackage{amssymb}
\usepackage{bm}
\usepackage{latexsym}
\usepackage[colorlinks=true,linkcolor=blue,citecolor=blue]{hyperref}
\usepackage{color}

\theoremstyle{plain}

\newcommand{\la}{\langle}
\newcommand{\ra}{\rangle}

\newcommand{\ti}{\tilde}
\newcommand{\ga}{\gamma}
\newcommand{\Ga}{\Gamma}
\newcommand{\ka}{\kappa}
\newcommand{\da}{\dagger}
\newcommand{\De}{\Delta}

\newcommand{\si}{\sigma}

\newcommand{\om}{\omega}
\newcommand{\Om}{\Omega}

\newcommand{\non}{\nonumber}
\newcommand{\pa}{\partial}

\def\pra#1{{ Phys.\ Rev. A\/} {\bf#1}}
\def\prb#1{{ Phys.\ Rev. B\/} {\bf#1}}

\def\prl#1{{ Phys.\ Rev.\ Lett.} {\bf#1}}
\def\pr#1{{ Phys.\ Rev.} {\bf#1}}

\def\rmp#1{{ Rev. \ Mod. \ Phys.} {\bf#1}}

\begin{document}

\title{Dynamics and control of a qubit in spin environment: an exact master equation study}

\author{Jun Jing$^{1,2}$ and Lian-Ao Wu$^{2,3}$}\thanks{Corresponding author: lianao.wu@ehu.es}

\affiliation{$^{1}$Institute of Atomic and Molecular Physics and Jilin Provincial Key Laboratory of Applied Atomic and Molecular Spectroscopy, Jilin University, Changchun 130012, Jilin, China \\ $^{2}$Department of Theoretical Physics and History of Science, The Basque Country University (EHU/UPV), PO Box 644, 48080 Bilbao, Spain \\ $^{3}$Ikerbasque, Basque Foundation for Science, 48011 Bilbao, Spain}

\date{\today}

\begin{abstract}
We present an exact master equation for a central spin-$1/2$ system coupled to a spin-$1/2$ bath in terms of hyperfine interaction, which provides a unified formalism for both free evolution and controlled dynamics of the central spin. The equation enables us to study controllability of dynamics processes with various degrees of non-Markovianity. We investigate the Overhauser's effect on decoherence dynamics of the central spin under different bath spectra and the system-bath coupling strengths. Nonperturbative leakage elimination approach is applied to the system to suppress decoherence in solid-state quantum information processing.
\end{abstract}

\maketitle

\section{Introduction} Understanding nonequilibrium dynamics of quantum systems interacting with a large number of uncontrollable degrees of freedom is a rapidly emerging topic, developed in various fields, such as quantum optics and quantum computation based on mesoscopic solid system~\cite{Nielson,Wiseman}. In spin-based systems, quantum computing proposals using quantum dots~\cite{Loss,Imamoglu} have led to intensive studies on coherent control of quantum degrees of freedom, in particular electron spins. Electron spin qubits in nuclear spin environments have two distinct merits: scalability and non-Markovianity, where the latter will be the focus of this study. The motion of electron spin in the nano-scale structures, where the quantum effects are important, is therefore crucial in fundamental research of open quantum systems~\cite{Breuer} and quantum control techniques or strategies~\cite{WuPRL03,WuPRA10}.

For solid spin-qubit systems, a variety of mechanisms have been identified to be responsible for the electron spin decoherence, such as the spin-orbital scattering with phonons, spectral diffusion due to dipolar interaction of nuclear spins, and the hyperfine interaction between electron spin and environmental nuclear spins. Low temperature and high magnetic field have been used to  successfully reduce decoherence due to these mechanisms but fail to suppress that due to hyperfine interaction. Under the hyperfine interaction, the nuclear spin bath is highly non-Markovian and the central spin has a long coherence time~\cite{Lukin1,Lukin2,Lukin3,Lukin4}. To understand these interesting observations an {\em exact} master equation for the central electron spin, as that for the spin-bath model~\cite{BreuerPRA}, is therefore desired.

The paper first derives an exact master equation for the central spin under the hyperfine interaction. This equation enables us to study dynamics of the central spin, and more interestingly to demonstrate the controlled dynamics by the nonperturbative leakage elimination operator (LEO) protocol. The nonperturbative LEO protocol was introduced recently~\cite{LEO} to suppress decoherence due to the presence of environment interference. The advantage of the protocol is that we can introduce a LEO to the system Hamiltonian, keeping the open system dynamics exactly solvable irrespective to the size of the system. Because of the exact solvability, we could understand correctly controllability of non-Markovianity and clarify possible confusion owing to the Markovian approximation.

\section{Exact master equation} Consider a central spin-$1/2$ embedded in a spin-$1/2$ bath through the hyperfine interaction:
\begin{eqnarray}\non
H_{\rm tot}&=&\om_0S_z+\sum_k\om_kI_k^z \\ \label{sH} &+& \sum_k\frac{A_k}{2}(S_+I_k^-+S_-I_k^+)
+\sum_kA_kS_zI_k^z,
\end{eqnarray} 
where the first two terms correspond to the Zeeman energy of the central and environmental spins, the operators $S$ and $I$ indicate the central spin and environmental spin modes, respectively, and $A_k$'s are the coupling strengths between the central spin and the $k$-th environmental spin. This is a typical spin-bath model characterizing physical entities such as atomic, molecular systems and artificial two-level systems.

In quantum dot systems, $\om_0$ is determined by the electron spin Zeeman effect of external magnetic field. The third and the fourth term of Eq.~(\ref{sH}) are termed as the flip-flop interaction and (longitudinal) Overhauser field, giving rise to inhomogeneous broadening and dephasing respectively. In the interaction picture with respect to $\om_0S_z+\sum_k(\om_k-A_k/2)I_k^z$, the total Hamiltonian becomes
\begin{equation}\label{sHI}
H^I_{\rm tot}=S_+(t)B^-(t)+S_-(t)B^+(t)+|1\ra\la1|B^z,
\end{equation}
where $B^\pm(t)=\sum_k\frac{A_k}{2}I_k^\pm e^{\pm i(\om_k-A_k/2)t}$, $B^z=\sum_kA_kI_k^z$ and $S_\pm(t)=S_\pm e^{\pm i\om_0t}$. Note that the last term in Eq.~(\ref{sHI}) is equivalent to $\sum_kS_zA_kI_k^z+\sum_k\frac{A_k}{2}I_k^z$. The total exciton number or angular momentum is conserved under the total Hamiltonian, such that one can work in one of invariant subspaces at a given exciton number. It is interesting to note that, the transversal hyperfine term $S_+(t)B^-(t)+h.c.$ results in off-resonant transitions between the system spin and environmental spins, while the longitudinal hyperfine term $|1\ra\la1|B^z$ provides additional contributions to the energy splitting.

In general an initial state $|\psi(0)\ra=C_0|0\ra|0\ra_E+c_0(0)|1\ra|0\ra_E+\sum_kc_k(0)|0\ra\si_k^+|0\ra_E$ evolves into the state $|\psi(t)\ra=C_0|0\ra|0\ra_E+c_0(t)|1\ra|0\ra_E+\sum_kc_k(t)|0\ra\si_k^+|0\ra_E$ at time $t$ according to the Schr\"{o}dinger equation represented in the single-exciton subspace,
\begin{eqnarray*}
\frac{d}{dt}c_0(t)&=&ihc_0(t)-i\sum_k\frac{A_k}{2}
e^{i(\om_0-\om_k+\frac{A_k}{2})t}c_k(t), \\
\frac{d}{dt}c_k(t)&=&-i\frac{A_k}{2}e^{-i(\om_0-\om_k+\frac{A_k}{2})t}c_0(t),
\end{eqnarray*}
where $h\equiv\sum_k\frac{A_k}{2}$. We now assume a fully polarized initial spin-bath state, $c_k(0)=0$, as done in Ref.~\cite{Lukin1} and evidenced by recent high polarization experiments in quantum hall edge states~\cite{DNP1} ($\sim85\%$) and a bias voltage in a ballistic quantum wire~\cite{DNP2} ($\sim94\%$). The coefficient $c_k(t)$ therefore satisfies
\begin{equation}\label{c0}
\frac{d}{dt}c_0(t)=ihc_0(t)-\int_0^tdsf(t-s)c_0(s),
\end{equation}
where the kernel function is given by a two-point correlation function of the reservoir $f(t-s)=\sum_k\left(\frac{A_k}{2}\right)^2e^{i(\om_0-\om_k+\frac{A_k}{2})(t-s)}=\la B^-(t)B^+(s)\ra_Ee^{i\om_0(t-s)}$, and the first term in Eq.~(\ref{c0}) is given by the Overhauser field in the total Hamiltonian~(\ref{sH}).

We now define a propagator that $c_0(t)=G(t)c_0(0)$, which does not depend on the initial condition due to the convex-linear characteristic in the decomposition of system initial state. Furthermore, we define $\ti{G}(t)\equiv G(t)e^{-iht}$ and can show that it satisfies
\begin{equation}\label{Gt}
\pa_t\ti{G}(t)=-\int_0^tds\ti{f}(t-s)\ti{G}(s),
\end{equation}
where $\ti{G}(0)=G(0)=1$ and $\ti{f}(t-s)\equiv f(t-s)e^{-ih(t-s)}$. To construct an exact master equation for the central spin in a time-convolutionless (TCL) form: $\pa_t\rho(t)=\mathcal{K}_{\rm TCL}(t)\rho(t)$, one can use an exact dynamical map $\Phi(t)$, which transforms the initial states into the states at time $t$: $\rho(t)=\Phi(t)\rho(0)$, i.e., $\mathcal{K}_{\rm TCL}(t)=\dot{\Phi}(t)\Phi^{-1}(t)$. After a straightforward derivation, we can obtain the exact TCL master equation
\begin{eqnarray}\non
\pa_t\rho(t)&=&-\frac{i}{2}S(t)[\si_+\si_-, \rho(t)] \\ \label{zero}
&+&\ga(t)\left[\si_-\rho(t)\si_+-\frac{1}{2}\{\si_+\si_-, \rho(t)\}\right],
\end{eqnarray}
where $S(t)=-2{\rm Im}(\dot{G}/G)$ and $\ga(t)=-2{\rm Re}(\dot{G}/G)$. Thus the stark-shift effect and the damping rate are determined by the imaginary and real parts of $\dot{G}/G$, respectively. It is evident that ${\rm Re}(\dot{G}/G)={\rm Re}(\dot{\ti{G}}/\ti{G})$, such that the damping rate is fully determined by Eq.~(\ref{Gt}). The fidelity $\mathcal{F}(t)\equiv\sqrt{\la\psi(0)|\rho(t)|\psi(0)\ra}$ is $|G(t)|$ or $|\ti{G}(t)|$.

\section{Decoherence dynamics of central spin} The dynamics of the exact TCL equation, parameterized by $S(t)$ and $\gamma(t)$, is given by the correlation functions $\ti{f}(t-s)$ determined by various environments. Now we first consider an exponential correlation function $\ti{f}(t-s)=\frac{\Ga\ga_0}{2}e^{-\ga_0|t-s|}$, where $\ga_0$ is usually understood as a measure of memory capacity or non-Markovianity of the environment. When $\ga_0$ approaches zero, the correlation function and $|\ti{G}(t)|$ become constants such that the corresponding environment memorizes the whole information between any two time moments $t$ and $s$. Contrarily, when it approaches infinity, the correlation function between any two time moments $t$ and $s$ becomes a delta function, such that the environment loses its capacity of memory. The solution of $\ti{G}(t)$ is
\begin{equation}\label{tGt}
\ti{G}(t)=e^{-\frac{\ga_0}{2}t}\left[\cosh\left(\frac{\ga_0\chi}{2}t\right)
+\frac{1}{\chi}\sinh\left(\frac{\ga_0\chi}{2}t\right)\right],
\end{equation}
where $\chi=\sqrt{1-2\Ga/\ga_0}$, and  $\ti{G}(t)=\mathcal{F}(t)$ in this special case. The coefficient $S(t)=0$ and the decay rate in the TCL equation~(\ref{zero}) is
\begin{equation*}
\ga(t)=\frac{2\Ga\sinh\left(\frac{\ga_0\chi}{2}t\right)}
{\chi\cosh\left(\frac{\ga_0\chi}{2}t\right)+\sinh\left(\frac{\ga_0\chi}{2}t\right)}.
\end{equation*}
An interesting example is that when $\ga_0=2\Ga$, $\ti{G}(t)=e^{-\Ga t}(1+\Ga t)$ and $\ga(t)=2\Ga^2t/(1+\Ga t)$. This means that the probability for the central electron spin staying at the upper level $|1\ra$ decays in the manner of $e^{-\ga_0t^2}$ in the short time limit, which is dramatically different from the normal exponential decay behavior for a two-level system embedded in a dissipative bosonic bath. It is interesting to note that the $t^2$ decay is similar to the Anti-Zeno effect shown in Ref.~\cite{WuS}.

\begin{figure}[htbp]
\centering
\includegraphics[width=3.0in]{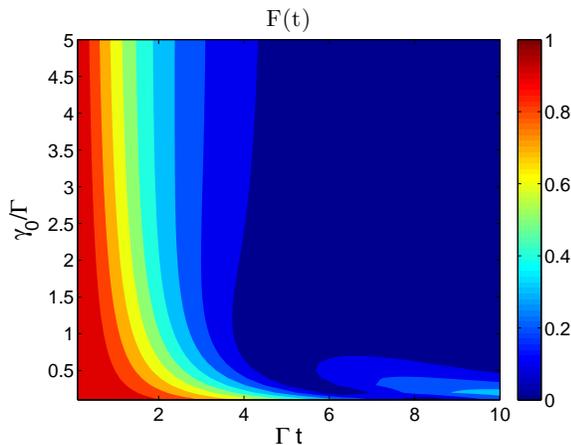}
\caption{$\mathcal{F}(t)=|\ti{G}(t)|$ in Eq.~(\ref{tGt}) as a function of dimensionless time $\Ga t$ and environmental memory parameter $\ga_0/\Ga$. }\label{Ffree}
\end{figure}

Figure~(\ref{Ffree}) shows dynamics of the fidelity $\mathcal{F}(t)$. The correlation function is a Lorentzian spectral density, and its half-width is $\ga_0$. Since realistic spin-bath spectra are distributed over a relatively narrow frequency regime, the Lorentzian function is not an accurate characterization but an approximation. Therefore the spectra must be truncated based on physical consideration, and in our case the truncation is set as $\ga_0/\Ga=5$, as shown in the figure~\ref{Ffree}. It is interesting to note that when $\ga_0/\Ga$ is below $0.5$, the fidelity revives, and information bounces from the spin bath back to the system. This is a typical non-Markovian dynamics. Another interesting phenomenon is that although in the short time limit, the decay speed of fidelity increases with $\ga_0$, whereas the moment when the system-spin completely decays is independent of a monotonic relation with the memory parameter. It is evident that when $\ga_0/\Ga<1$, the fidelity vanishes at a time which decreases with $\ga_0$. Moreover the fidelity reaches its maximum around $\ga_0/\Ga=1$, and then declines asymptotically to a steady value when $\ga_0/\Ga>1$.

We now consider the case when $A_k\approx\mathcal{A}/N$ and $\om_k\approx\om$, where $N$ is the number of environmental spins. For this model, we have $h=\mathcal{A}/2$, and $\ti{f}(t-s)=\frac{\mathcal{A}^2}{4N}e^{i\Om(t-s)}$, where $\Om=\om_0-\om-\mathcal{A}/2(1-1/N)$. Inserting it into Eq.~(\ref{Gt}), one can find the solution
\begin{equation}
\ti{G}(t)=\frac{\De-\Om}{2\De}e^{i(\Om+\De)t/2}+\frac{\De+\Om}{2\De}e^{i(\Om-\De)t/2},
\end{equation}
where $\De=\sqrt{\Om^2+\mathcal{A}^2/N}$. The system will periodically come back to its initial state when the absolute value of $|c_0(t)|=|\ti{G}(t)|=1$, i.e., $\De t=2n\pi$ with $n$ arbitrary integer. This ``box'' model describes that the central spin coherence can be reversibly transferred into a collective state of the surrounding environmental spins, and allows for exploiting the environmental spins to store quantum information in the central spin. The leading decoherence mechanism for the stored state is bath-spin diffusion, with dephasing rates in the kHz domain. Techniques similar to spin-echo~\cite{SE,Viola} can be used to mitigate its effect.

Solid (semiconductor) spin-based qubits are promising candidates for quantum computation because of their scalability. The fundamental single-qubit gates have been demonstrated for GaAs-based spin qubits~\cite{Hanson}. The typical data of the III-V semiconductor compounds for quantum dot~\cite{Hanson} shows that $|\om_0|\approx|\mathcal{A}|\approx(10^2\sim10^3)|\om_k|\approx(10^4\sim10^6)|A_k|$, implying that the ``box'' model is a reasonable idealization. The inherent error caused by the hyperfine coupling therefore features with a non-Markovian character and then can be controlled to some extent. The Overhauser effect is determined by signs of $A_k$ or $h$ relative to $\om_0$. If their signs are the same, $\Om$ will be reduced. This is equivalent to reduce the energy splitting of the central spin and makes it more fragile to the decoherence induced by flip-flop term. Otherwise, the Overhauser field could naturally protect the system coherence.

Let us now consider a more realistic situation where $\Om$ and $A$ are $k$-dependent, i.e., $\Om_k=\om_0-\om_k-A_k/2$, and then $\ti{f}(t,s)=\sum_k\frac{A_k^2}{4}e^{i\Om_k(t-s)}$. As discussed in the ``box'' model, it is reasonable to assume that $A_k$ satisfy a Gaussian distribution characterized by the mean value $\mu$ and the variance $\nu^2$, where $\mu$ and $\nu$ are in the same order of $|\mathcal{A}|/\sqrt{N}$. Since in reality $\om_0$ and $\mathcal{A}$ are much larger than $\om_k$, $\Om_k$ can be approximated to a continuous variable centering the average value $\mathcal{A}$. Therefore when $\mathcal{A}\om_0>0$, the effective correlation function of the spin-bath can be written as
\begin{equation}\label{fts}
\ti{f}(t-s)\approx\frac{\mathcal{A}}{2\sqrt{N}}e^{-\frac{\nu^2}{2}(t-s)^2
+i\frac{\mathcal{A}}{2}(t-s)},
\end{equation}
which is insensitive to the mean frequency of the bath-spins since most of $A_k$'s are much less than $\mathcal{A}$ and $\mu\ll\mathcal{A}$. Similarly, $\ti{f}(t-s)$ and $\mathcal{F}(t)=|\ti{G}(t)|$, which can be numerically obtained by inserting Eq.~(\ref{fts}) into Eq.~(\ref{Gt}), are also insensitive to $\nu$ because $\nu\ll\mathcal{A}$.

\begin{figure}[htbp]
\centering
\includegraphics[width=3.0in]{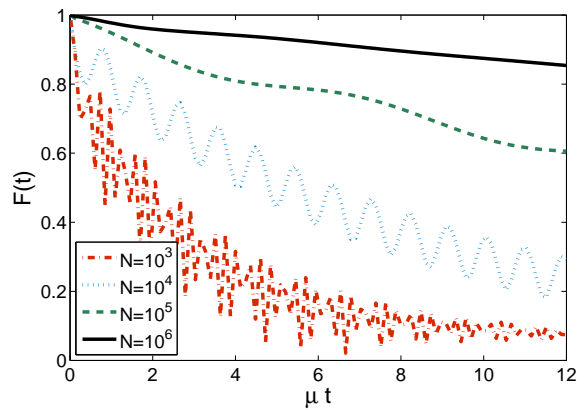}
\caption{ $\mathcal{F}(t)=|\ti{G}(t)|$ as a function of the dimensionless time $\mu t$, where $\mu$ is the average frequency of $\om_k$, and $N$ is the number of bath-spins.}\label{Ffree2}
\end{figure}

Figure.~(\ref{Ffree2}) demonstrate $\mathcal{F}(t)$ as a function of the number of environmental spins $N$, we set $\nu=0.5\mu$ and $\mathcal{A}=10^2\mu$. It is interesting to compare the decoherence patterns for different $N$. When $N=10^4$, the fidelity decays with a strong fluctuation, meaning that exchange of the quantum information and energy between the system and bath spins is significant. When $N$ increases, the amplitude of fluctuation shrinks and the decay rate becomes smaller and smaller. As such the coherence of the central spin remains robust for large $N$. 

\section{Leakage elimination in the system spin} We can apply an open-loop control technique using Leakage Elimination Operator to strengthen the robustness of the central spin system~\cite{LEO}. Specifically, LEO for this system is $R_L=\sigma_z$ and  the corresponding Hamiltonian is $H_L=r(t) \sigma_z$, where $r(t)$ is a time-dependent function. A distinguished character for $R_L$ is, as shown in the pervious and present studies, that $r(t)$ can be almost arbitrarily chosen, even chaotic or noisy signals.

\begin{figure}[htbp]
\centering
\includegraphics[width=3.0in]{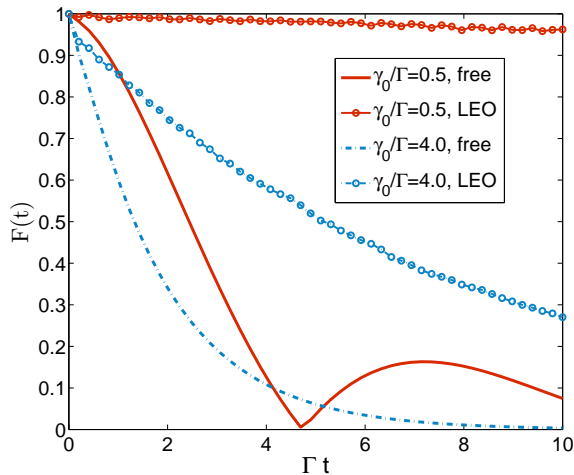}
\caption{The free and controlled dynamics of $\mathcal{F}(t)=|\ti{G}(t)|$ as a function of dimensionless time $\Ga t$ with different environmental memory parameter $\ga_0$. In the control of LEO protocol, we apply a random pulse sequence, whose period and duration time are $\tau=0.02\Ga t$ and $\ka=0.01\Ga t$, respectively, and the average strength is $\Psi=0.2\Ga$.}\label{Fcont}
\end{figure}

By using the fast bang-bang pulses circuit and $\{R_L, S_-(t)B^\da(t)+h.c.\}=0$ and $[R_L, |1\ra\la1|B^z]$, it follows that $\lim_{m\rightarrow\infty}(e^{-i\frac{H^I_{\rm tot}t}{m}}R^\da_L
e^{-i\frac{H^I_{\rm tot}t}{m}}R_L)^m=e^{-i|1\ra\la1|B^zt}$, where we apply the Trotter formula that the evolution operator $U(t)=\lim_{m\rightarrow\infty}(e^{-i\frac{H^I_{\rm tot}t}{m}})^m$. This result holds to the order of $t^2$ only and is an idealization or perturbation. In the nonperturbative case, the presence of $H_L$ in the total Hamiltonian will modify the kernel $\ti{f}(t-s)$ into
\begin{equation}\label{g}
g(t-s)=\ti{f}(t-s)e^{-i\int_s^tds'r(s')}=\ti{f}(t-s)e^{-iR(t-s)},
\end{equation}
where $R\equiv\frac{\int_s^tds'r(s')}{t-s}$. According to the Riemann-Lebesgue lemma, if $|\ti{f}(t-s)\ti{G}(s)|$ is finite, which is evident, then the integral in Eq.~(\ref{Gt}) approaches zero as $R\rightarrow\infty$. Therefore when $\ti{R}(t)\equiv\int_0^tdsr(s)$ is sufficient large, one can use LEO to suppress the decoherence of the system spin. Consequently, the effectiveness of LEOs depends solely on the integral $R$ in the time domain but not the details of $r(t)$.

Figure~(\ref{Fcont}) demonstrates the free and controlled dynamics of $\mathcal{F}(t)$. In the LEO control protocol, we use an equiv-distant rectangular pulse sequence, whose intensity is random described by ${\rm rand}(t)\Psi/\ka$ in the duration time $\ka$ and is zero in the dark time for each period $\tau$, where ${\rm rand}(t)$ is a random number between zero and one and $\Psi$ is the average strength. The figure shows that LEO control is excellent when it is applied in a strong non-Markovian regime with a longer memory time characterized by $1/\ga_0$. Therefore LEO is a reliable tool to decouple the system spin from the influence of spin bath. Our formalism clarifies a fact that {\em an ideal Markovian process cannot be controlled} because in the Markovian limit where $\ti{f}(t-s) \propto \delta(t-s)$, the control factor $e^{-i\int_s^tds'r(s')}$ in Eq.~(\ref{g}) becomes invalid. In other words, a Markovian dynamics cannot be affected by external controlled fields.

\section{Conclusion} We have proposed an exact master equation for a central spin coupled to a spin bath. We analyze its decoherence dynamics in the presence of a Lorenzian and a Gaussian spectral density functions, respectively.  We have found that the Overhauser field may help to suppress the decoherence process of the system spin, and the system evolution may be periodic in the ``box'' approximation. Leakage elimination operator is applied to the system, and we show that the operator can protect coherence of the central spin. Although our master equation is obtained based on the full polarization assumption, yet it still provides a convenient way to capture the non-Markovian feature of the spin bath. The control technology hinted by the master equation can be used in spin-based quantum setups, in which the non-Markovianity (the long-time memory capability) is beneficial to maintain coherence of the central spin system.

\section*{Acknowledgments} We acknowledge grant support from the Basque Government (grant IT472-10), the Spanish MICINN (No. FIS2012-36673-C03-03), the National Science Foundation of China Nos. 11175110, 11575071 and Science and Technology Development Program of Jilin Province of China (20150519021JH).

\end{document}